# Dissolution of a two-component drop onto macrophase due to surface tension effect

Alexey Kabalnov, Ink Splat Company, kabalnov@inkjet3d.com

## Abstract

Additives of sparingly soluble components are known to slow down or completely inhibit Ostwald ripening in dispersed systems. In this paper series, our earlier model of the stabilization against Ostwald ripening is revisited and extended over the whole range of compositions, molar volumes of components, and their activity coefficients. In the first paper, a simpler problem, the dissolution of a two-component drop under the action of excess Laplace pressure inside is analyzed.
Three stages of dissolution are identified. In the first stage, called pre-lock-in, the concentration of the poorly soluble component undergoes a quick increase, and the system enters the lock-in state, in which the Laplace pressure effect on the chemical potential of the more soluble component is nearly completely counterbalanced by the Raoult effect. After this, the dissolution kinetics slows down and enters a steady state. In the process, the concentration of the sparingly soluble component continues to increase, first slowly and then more rapidly in the very end of the particle lifetime; this latter stage is called the 'late lock-in'. Despite all those variations, if the initial concentration of the poorly soluble component is above a certain threshold, the dissolution kinetics nearly follows the classical cubic law. An improved extrapolatory equation for the rate of dissolution is proposed that covers the whole formulation range and represents an extension over our previous formula [2].

## Introduction

Since the pioneering work of Higuchi and Misra [1], it was realized that an insoluble additive to the dispersed phase can slow down or eliminate Ostwald ripening in dispersions. In our earlier work, possible mechanisms of such stabilization were explored and the first qualitative theory developed [2]. Webster and Cates extended the model for the case of a completely insoluble additive [3]. However, the case when the additive has a finite solubility in the medium has not been sufficiently studied. This is the topic of this paper series.
Before jumping into the Ostwald ripening theory, a simpler problem, the dissolution of a two-component droplet under the action of excess Laplace pressure is analyzed. This model in general does not require the solution of the components to be ideal, and covers the whole range of compositions, within the molar fraction range from 0 to 1.

The paper has the following outline. We start with setting up the diffusion equations for the problem in general case. Then we analyze a particular case when one of the components is substantially less soluble in the medium than the other; the analysis is done first on the level of chemical potentials and then solubilities of the components. A particular diffusion mode is shown to develop that is called the lock-in mode; in this mode, the diffusion of the less soluble component controls the rate of the overall process. The model allows to predict the dissolution rate of the particle in general case, including the case of non-ideal solutions, but utilizes the assumption of smallness of the relative variations in the particle composition during the dissolution. In the next section, we review the conditions for such approximation to hold. The numerical simulations to validate the model will follow. In the next section, we explore in more detail the rate of 'bleeding' of the poorly soluble component from the particle, and the validity of the extrapolatory equation that connects the different dissolution modes. In conclusion, we briefly discuss the connection of the model to Ostwald ripening, which will be mostly addressed in the follow-up paper.

**Dissolution of a two-component drop under excess Laplace pressure**

Consider a spherical liquid drop formed by two completely miscible components, 1 and 2. Both components are soluble in the medium, with the solubilities for the individual components $C^*_{\infty 2}$and $C^*_{\infty 2}$ respectively. The infinity subscript means the solubility of the macrophase, that is, the infinite radius of curvature. Note that those the values are of the pure components; the solubilities from their mixture are different, as the components reduce the solubility of each other per classical solution theory; this analysis will be the partially the object of this study.
We assume that the second component is substantially less soluble in the medium by itself; $C^*_{\infty 2} << C^*_{\infty 1}$. Let the molar fractions of the components in the particle be $x_1$ and $x_2$, respectively; at time zero, the values are $x_{01}$ and $x_{02}$ . We use the subscript 0 to indicate the time 0, that is, the composition of the drop before the mass transfer has started. The droplet is in a vicinity of a macrophase that initially has the same composition as the drop, and a zero curvature. The medium surrounding the particle is in equilibrium with the macrophase, except for the adjacent area where the concentration gradient develops. Over the time, the drop gradually dissolves because of the excess Laplace pressure; during this process its composition also changes, however the composition of the macrophase is assumed to be constant (Figure 1).

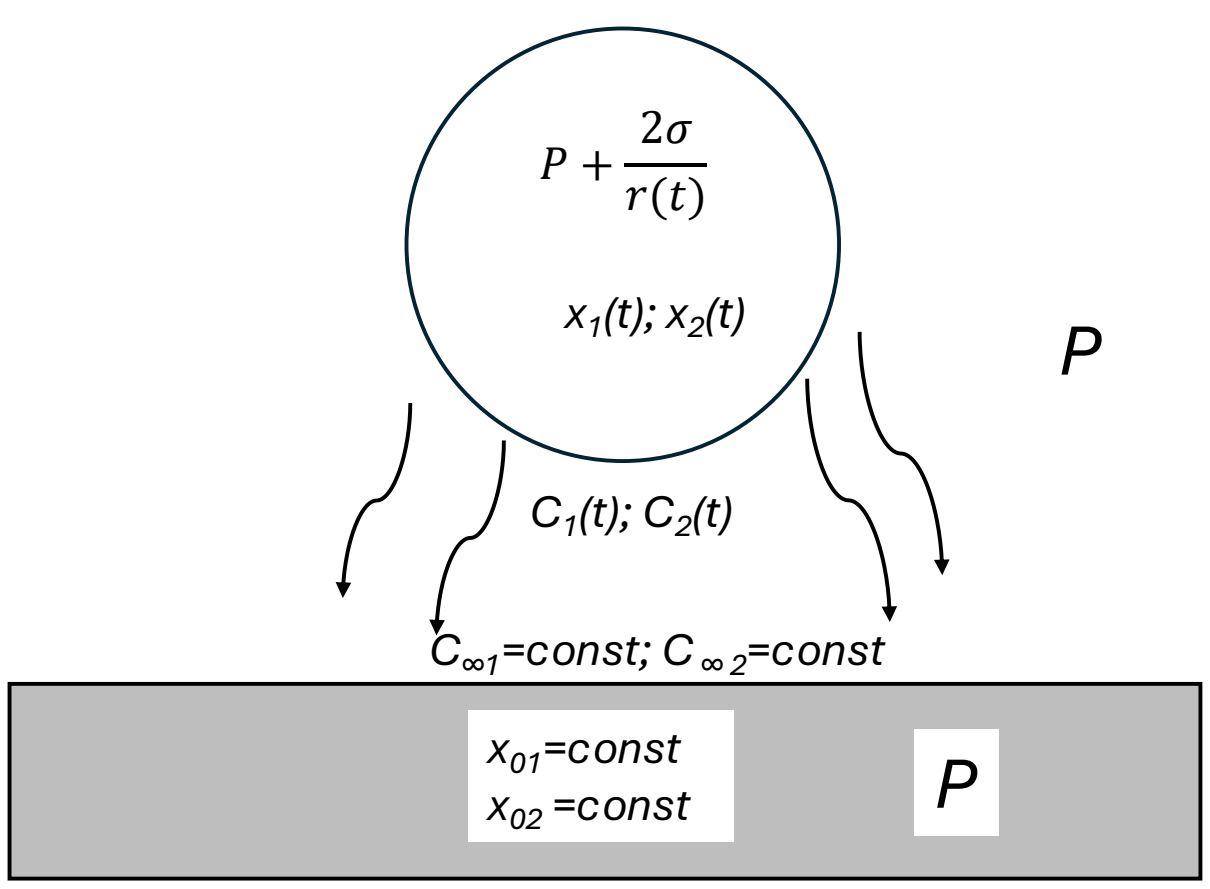

Figure 1. Cartoon illustrating the dissolution problem setup

For the dissolution from a spherical particle, the steady state mass flows, $J_1$ and $J_2$ of the components are equal to [4]:

$$J_1 = 4\pi r D_1 \Delta C_1 \tag{1}$$

$$J_2 = 4\pi r D_2 \Delta C_2 \tag{2}$$

Here $r$ is the particle radius and $D_1$ and $D_2$ are the molecular diffusion coefficients. Finally, the concentration differences:

$$\Delta C_1 = C_1(t) - C_{\infty 1} \tag{3}$$
$$\Delta C_2 = C_2(t) - C_{\infty 2} \tag{4}$$

are the differences between the concentrations at the surface of the particle, and in the infinity, at the macrophase, respectively. We consider the concentration at the particle to be time dependent, as both the particle composition and the Laplace pressure change with time; on the other hand, the concentration at the macrophase is considered to be constant, and to correspond to the initial particle composition $x_{01}$ and $x_{02}$, and zero Laplace pressure.
The difference in chemical potentials $\Delta\mu_{1,2}$ between the drop and the macrophase are shown below:

$$\Delta\mu_1 = \mu_1(x_1, \Delta p) - \mu_1(x_{01}, \Delta p = 0) \tag{5}$$
$$\Delta\mu_2 = \mu_2(x_2, \Delta p) - \mu_2(x_{02}\ \Delta p = 0) \tag{6}$$

At time zero, all the difference comes only from the excess Laplace pressure; however, with time, there is also a contribution from the changes in $x_1$ and $x_2$, which we call the Raoult effect contribution. The equilibrium concentrations at the surface of the drop can be evaluated from these values of the excess chemical potentials:

$$RT ln\left(\frac{C_1}{C_{\infty 1}}\right) = \Delta\mu_1 \quad (7)$$

$$RT ln\left(\frac{C_2}{C_{\infty 2}}\right) = \Delta\mu_2 \quad (8)$$

As the excess chemical potentials are much less than *RT* in the range of Laplace pressures of interest, the exponents (or logarithms) can be expanded in series and

$$C_1 = C_{\infty 1}\left(1 + \frac{\Delta\mu_1}{RT}\right) \quad (9)$$

$$C_2 = C_{\infty 2}\left(1 + \frac{\Delta\mu_2}{RT}\right) \quad (10)$$

and

$$\Delta C_1 = C_{\infty 1}\frac{\Delta\mu_1}{RT} \quad (11)$$

$$\Delta C_2 = C_{\infty 2}\frac{\Delta\mu_2}{RT} \quad (12)$$

The values of the excess chemical potentials change during the dissolution of the particle, and this paper will be evaluating them as a function of time. Note again that $C_{\infty 1}$ and $C_{\infty 2}$ are constant, but different from $C^*_{\infty 1}$and $C^*_{\infty 2}$ because of the Raoult effect.

**Lock-In Approximation**

We start with the case when the Raoult effects on dissolution fully dominate over the capillary effects, in a sense of the paper [2]. The exact numerical criteria for this will be defined later; we just use it now in some loose qualitive terms. Consider a two-component droplet. The capillary pressure inside the particle makes the components to dissolve and diffuse into the macrophase. The more soluble Component 1 quickly diffuses out whereas Component 2 lags in diffusion. This continues until the Raoult effect due to the increased concentration of the Component 2 in the particle completely counterbalances the capillary driving force. Then, the process becomes controlled by the slow 'bleeding' of Component 2 from the particle. We will call this a 'lock-in state' and will analyze its kinetic below.

We consider both chemical potentials as a function of $x_2$ as the independent variable, and the excess pressure $\Delta p$ of Laplace origin. As we are interested in very small changes in composition *and* pressure, we can expand the chemical potentials in series:

$$\Delta\mu_1 = \frac{\partial\mu_1}{\partial x_2}\Delta x_2 + \frac{\partial\mu_1}{\partial p}\Delta p \qquad (13)$$

$$\Delta\mu_2 = \frac{\partial\mu_2}{\partial x_2}\Delta x_2 + \frac{\partial\mu_2}{\partial p}\Delta p \qquad (14)$$

Here $\Delta x_2 = x_2 - x_{02}$.

Considering that $\frac{\partial\mu_1}{\partial p} = V_{m1}; \frac{\partial\mu_2}{\partial p} = V_{m2}$ and $\Delta p = 2\sigma/r$, we conclude:

$$\Delta\mu_1 = \frac{\partial\mu_1}{\partial x_2}\Delta x_2 + \frac{\alpha_1}{r}RT \qquad (15)$$

$$\Delta\mu_2 = \frac{\partial\mu_2}{\partial x_2}\Delta x_2 + \frac{\alpha_2}{r}RT \qquad (16)$$

where

$$\frac{2\sigma V_{m1}}{rRT} \equiv \frac{\alpha_1}{r} \qquad (17)$$

$$\frac{2\sigma V_{m2}}{rRT} \equiv \frac{\alpha_2}{r} \qquad (18)$$

We call the first term of equations (15, 16) the Raoult term, and the second the Laplace term. For Component 1, Laplace and Raoult effects act in opposite directions (as the Raoult effect is negative) and counterbalance each other to nearly zero net effect, whereas for the Component 2 they are of the same sign and add together. To proceed further with the analysis, we assume that the difference between the Raoult and Laplace terms for the first component in the lock-in is nearly 0, that is, the difference between the terms is much smaller than the terms themselves[1].

We then add this condition to the chemical potential balance:

$$\Delta\mu_1 = 0 \qquad (19)$$

$$\Delta\mu_2 > 0 \qquad (20)$$

---

[1] A similar approach to estimate the rates in chemical kinetics is known as Bodenstein Steady State Method [5], where the rate of change of the concentration of a reactive intermediate is equated to zero, as the rate of its production is nearly equal to the rate of consumption.

Finally, we can apply the Gibbs-Duhem equation to the chemical potential increments. Luckily, Gibbs-Duhem equation enables to relate $\Delta\mu_1$ and $\Delta\mu_2$ for the general case, no matter if the solution of Components 1 and 2 in each other is ideal or not.

$$\frac{\partial\mu_1}{\partial x_2}x_1 + \frac{\partial\mu_2}{\partial x_2}x_2 = 0 \quad (21)$$

After some algebra we can get rid of the partial derivatives of the chemical potentials completely and conclude:

$$\Delta\mu_2 = \frac{2\sigma}{r}\frac{(V_{m1}x_1 + V_{m2}x_2)}{x_2} \quad (22)$$

The term $V_{m1}x_1 + V_{m2}x_2$ can be interpreted as the molar-weighted-average molar volume of the components, $\bar{V}_m$ . Equation (22) is similar to the Kelvin equation for a single component drop, with the difference that the molar volume is replaced by the combination $\bar{V}_m/x_2$. We now nearly ready to evaluate $\Delta C_2$ in the equation for the mass flow; before doing this however we need to evaluate $C_{\infty 2}$ . As based on the equality of the chemical potentials between the phases in equilibrium for the Component 2 we conclude:

$$\Delta\mu_2 = RTln\frac{C_2^*}{C_{\infty 2}} = RTln\frac{1}{x_2\gamma_2(x_2)} \quad (23)$$

Here on the left is the difference of the chemical potentials in the solution in the medium, and on the right, in the solution of the dispersed phase. We have 1 in the numerator on the right-hand side because for the pure Component 2, $x^*_2 = 1$ and the activity coefficient $\gamma_2$=1. The mass flow of Component 2 in g/cm$^3$ from a spherical particle in a steady state, is then equal to:

$$J_2 = -\frac{8\pi\sigma\bar{V}_m(x_2)D_2C_2^*\gamma_2(x_2)}{RT} \quad (24)$$

The minus sign at the flux indicates that the particle is dissolving. From this point on, we use the dimensionless units for the concentration, which we obtain by dividing the concentration expressed in g/cm$^3$ to the density of the second component, expressed in g/cm$^3$. The mass flow then becomes the volume flow, expressed in cm$^3$/s. To determine the change of the particle volume with time, we note that the system is in the locked-in state, and the molar fractions $x_1$ and $x_2$ remain nearly constant at their initial $x_{01}$ and $x_{02}$ values, as the Raoult factor dominates over the Laplace factor. So are the volume fractions $\phi_1$ and $\phi_2$; the limits to validity of this approximation are an important topic that will be discussed in the next section. Within this approximation, $\bar{V}_m(x_2) \approx \bar{V}_m(x_{02}) \equiv \bar{V}_m$ and $\gamma_2(x_2) \approx \gamma_2(x_{02}) \equiv \gamma_2$ . In order to keep the volume fractions constant, the overall volume should decrease proportionally to the partial volume of each component[2]. That is, if

$$\frac{d\phi_2}{dt} = \frac{\frac{dV_2}{dt}V - \frac{dV}{dt}V_2}{V^2} = 0 \qquad (25)$$

then:

$$\frac{dV_2}{dt} = J_2 = \frac{dV}{dt}\phi_{02} \qquad (26)$$

and

$$\frac{dV}{dt} = \frac{J_2}{\phi_{02}} = -\frac{8\pi\sigma\bar{V}_m C_2^* \gamma_2 D_2}{\phi_{02} RT} \qquad (27)$$

Note that if the first and second components would be dissolving alone, as a single component, the rates would have been:

$$\frac{dV}{dt} = -\frac{8\pi\sigma V_{m1} C_1^* D_1}{RT} \qquad (28)$$

and

$$\frac{dV}{dt} = -\frac{8\pi\sigma V_{m2} C_2^* D_2}{RT} \qquad (29)$$

These rates should bracket the equation (27) for the mixture. One of the challenges is to connect these three kinetic regimes into a single equation. If such an equation would be found, it will cover the whole range of compositions, from $\phi_{02}$ = 0 to 1. We also aspire to include the possible deviations of the solution from ideality into this equation.

[2] the volume of mixing effect is neglected throughout of this paper

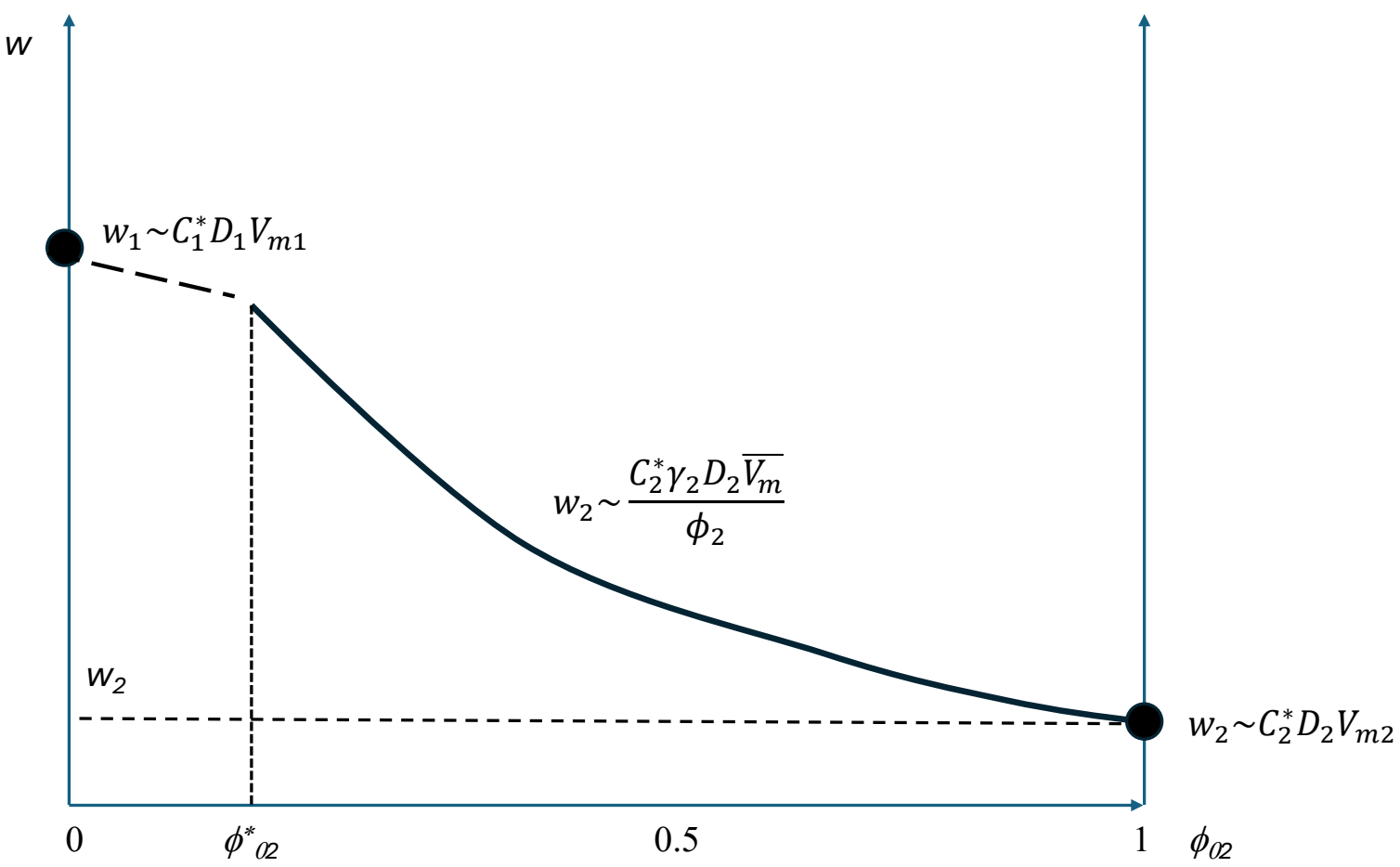

Figure 2. Cartoon illustrating dissolution of the drop over the whole range of compositions of $\phi_{02}$ from 0 to 1. The 'exactly known' ranges are shown in bold; the extrapolated range is shown as a dashed line.

Figure 2 illustrates the dissolution rates in the 2-component system. Equation (27) is covering the whole range of compositions, including the case of pure Component 2, as in this state $\overline{V}_m = V_{m2}$ and $\phi_{02} = 1$. It starts to break down on the other end of the curve at the value of $\phi_{02} = \phi_{02}^*$ when the rate as predicted by eqn (27) reaches the value of $w_1$. One of the ways to cover the dashed line region it is to combine the rates into an extrapolatory equation:

$$\frac{\overline{V}_m}{w} = \frac{\phi_{01} V_{m1}}{w_1 \gamma_1} + \frac{\phi_{02} V_{m2}}{w_2 \gamma_2} \qquad (30)$$

or, equivalently,

$$w = \frac{8\pi\sigma\overline{V}_m}{RT}\left(\frac{\phi_{01}}{C_{01}^* D_1 \gamma_1} + \frac{\phi_{02}}{C_{02}^* D_2 \gamma_2}\right)^{-1} \qquad (31)$$

If $\phi_{02}$ is large, the second term in paratheses is dominating, and Eqn 27 is recovered; conversely, if $\phi_{02} << 1$ and $\phi_{01}$ is close to 1 (which also means that $\gamma_1$ is close to 1), Eqn 28 is recovered. The transition between these two modes happens at $\phi_{02}^*$ such that

$$\frac{1-\phi_{02}^*}{\phi_{02}^*} \sim \frac{C_{01}^* D_1 \gamma_1}{C_{02}^* D_2 \gamma_2} \qquad (32)$$

We will call the ratio:

$$L_2 \equiv \frac{\phi_{02} C_{01}^* D_1 \gamma_1}{(1-\phi_{02}) C_{02}^* D_2 \gamma_2} \qquad (33)$$

the Second Lock-in Number; the significance of this ratio will be discussed in the following section. If the solution of the components in each other is ideal, the equation (31) reduces to:

$$w = \frac{8\pi\sigma\overline{V}_m}{RT}\left(\frac{\phi_{01}}{C_{01}^* D_1} + \frac{\phi_{02}}{C_{02}^* D_2}\right)^{-1} \qquad (34)$$

Finally, if the molar volumes of the components are nearly identical, we get:

$$w = \frac{8\pi\sigma\overline{V}_m}{RT}\left(\frac{\phi_{01}}{C_{01}^* D_1} + \frac{\phi_{02}}{C_{02}^* D_2}\right)^{-1} \qquad (35)$$

which reduces to the extrapolatory equation of our previous paper [2]:

$$w = \left(\frac{\phi_{01}}{w_1} + \frac{\phi_{02}}{w_2}\right)^{-1} \qquad (36)$$

The result of this paper is therefore an extension of our previous paper's result, and accounts for the difference in the molar volume of the components and the fact that the mutual solution of the components can be non-ideal. Whereas eqn 27 is 'exact', eqn 31 is extrapolatory and requires some numerical validation. In the next section we will be conducting numerical calculations to show the limits of accuracy of eqn 31.

**Limitations of the model: assumption of constancy of the droplet composition during dissolution**

The major simplification of the model relies on the approximation of eqn 26, which assumes that the composition of the particle stays approximately constant during its lifetime:

$$\frac{\Delta\phi_2(t)}{\phi_{02}} << 1 \qquad (37)$$

It allows to reconstruct the changes in the particle size as based on the mass flow of the poorly soluble component. In this section we will review, what conditions must be met for this to hold. When entering in a lock-in state, the excess chemical potential of Component

1 must become nearly zero as the Laplace and Raoult terms compensate each other. For ideal solutions it means:

$$\frac{\Delta\mu_1}{RT} = \frac{\alpha_1}{r} + ln\frac{x_1}{x_{01}} \approx 0 \qquad (38)$$

Thus, the molar fraction of the first component needs first to decrease in a stepwise fashion, $\Delta x_1 = -\Delta x_2 < 0$ in response to Kelvin effect. Expanding the logarithm in series and neglecting the higher order terms, one concludes that the relative change of the molar fraction while entering the lock-in is equal to:

$$\frac{\Delta\phi_2}{\phi_{02}} \approx \frac{\Delta x_2}{x_{02}} = \frac{\alpha_1(1-x_{02})}{r_0 x_{02}} \qquad (39)$$

If this ratio is required to be small, we need:

$$L_1 \equiv \frac{x_{02} r_0}{\alpha_1(1-x_{02})} >> 1 \qquad (40)$$

Here we introduce the dimensionless group $L_1$, which will be called the First Lock-in Number. It characterizes the magnitude of the relative change in concentration of the less soluble component as the system enters the lock in. However, we note that these are not all the changes we need to account for. After the lock-in, the molar fraction of the second component continues to slowly increase with time as the radius of the drop continues to diminish:

$$x_2(t) = \frac{\alpha_1(1-x_{02})}{r(t)} + x_{02} \qquad (41)$$

A schematic graph illustrating equation (41) is shown in Figure 3.

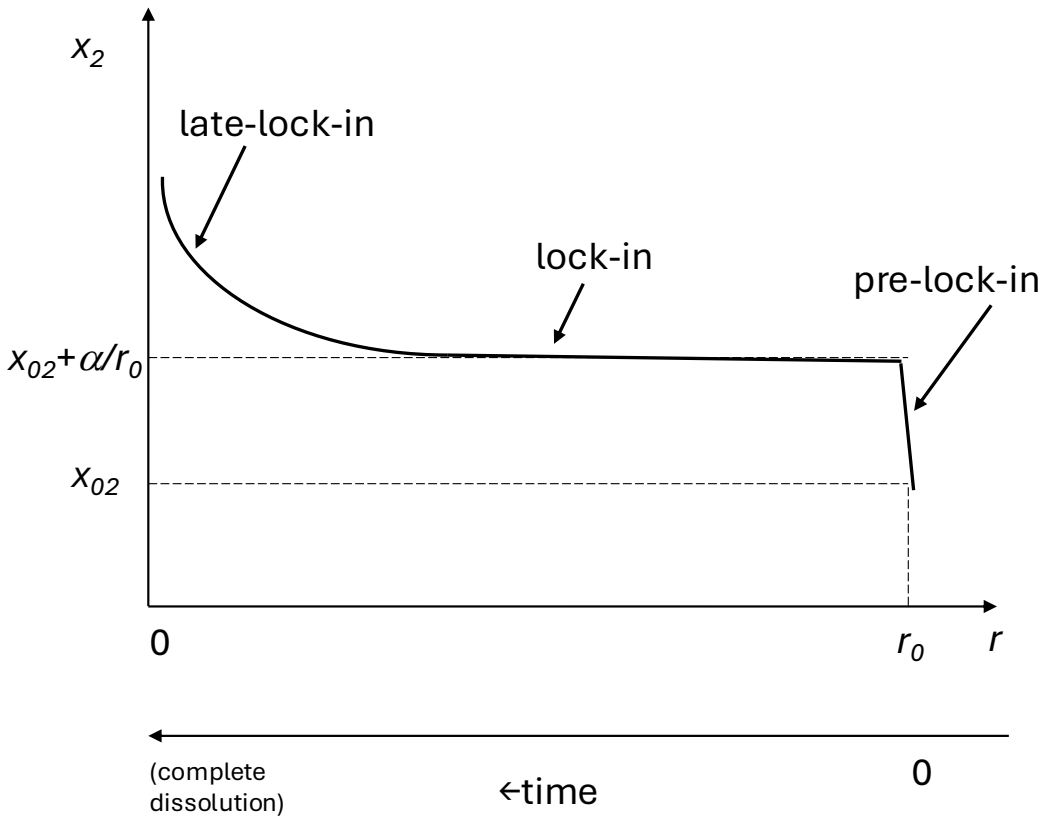

Figure 3. Cartoon, illustrating the change in the molar fraction of the second component over the life of the particle.

Eventually, the radius becomes such the condition is no longer met as

$$\frac{\Delta x_2}{x_{02}} = \frac{\alpha_1(1-x_{02})}{r(t)x_{02}} \approx 1 \qquad (42)$$

We call the stage of kinetics the 'late lock-in stage'. Below, we will evaluate its effect on the overall dissolution kinetics.

To do this, we note that Eqn (24) provides the value of the volume flux for Component 2 in general case. No longer assuming that $\phi_2 = x_2$ is constant, as it was in the previous section, we note:

$$J_2 = \frac{dV}{dt}x_2 + \frac{dx_2}{dt}V \qquad (43)$$

After some algebra, we arrive to the following law for the dissolution kinetics:

$$\Omega_2 = \left(4\pi x_{02}r^2 + \frac{8\pi\alpha_1}{3}r\right)\frac{dr}{dt} \qquad (44)$$

Here $\Omega_2 = -\frac{8\pi\sigma V_m D_2 C_2^*}{RT} = const$ is the dissolution rate of Eqn 24. Again, the minus sign indicates the dissolution. After integration by time, one gets:

$$-\Omega_2 t = x_{02}\frac{4\pi(r_0^3-r^3)}{3} + \frac{4\pi\alpha_1(r_0^2-r^2)}{3} \qquad (45)$$

The first term in the rhs is the cubic term that we had before in eqn (27). Accounting for the changes in the particle composition added the quadratic term to the model. The cubic term dominates at the early stage of the life of the particle, after which the quadratic term becomes progressively significant.
There are multiple ways to check the significance of this correction to the dissolution kinetics. One of the ways is the comparison of the times to complete dissolution of the particles by these two models. As follows from eqn (45), the ratio of is equal to:

$$\frac{\tau_2}{\tau_1} \approx 1 + \frac{1}{L_1} \tag{46}$$

The high $L_1$ numbers >10 in general indicate the negligibility of the quadratic term correction. Another way is to look at this problem is to evaluate the linearity of the volume versus time plots at the different values of $L_1$; it will be done in the next section. On the physical side, the prediction of the model is that, although the quadratic term does provide a correction to the dissolution kinetics, the correction is small if values of $L_1$ are large. Note that at the late lock in, the relative changes in concentration are significant, but its effect on the particle lifetime is small.

**Pre-lock-in, lock-in and late lock-in stages: numerical simulations**

To further analyze the limits of the validity of the lock-in assumption, we will conduct the numerical calculations of the particle dissolution, no longer assuming that $\Delta\mu_1 = 0$ and solving the differential equations directly. For this, we will be considering the solution of the Components 1 and 2 in each other to be ideal. The equations for the excess chemical potentials and concentrations are shown below:

$$\frac{\Delta\mu_1}{RT} = ln\left(\frac{x_1}{x_{01}}\right) + \frac{\alpha_1}{r} = ln\left(\frac{C_1}{C_{\infty1}}\right) \tag{47}$$

$$\frac{\Delta\mu_2}{RT} = ln\left(\frac{x_2}{x_{02}}\right) + \frac{\alpha_2}{r} = ln\left(\frac{C_2}{C_{\infty2}}\right) \tag{48}$$

$$\Delta C_1 = C_{\infty1}\frac{x_1}{x_{01}}exp\left(\frac{\alpha_1}{r}\right) - C_{\infty1} \approx C^*_{\infty1}x_{01}\left[\frac{x_1}{x_{01}}\left(1 + \frac{\alpha_1}{r}\right) - 1\right] \tag{49}$$

$$\Delta C_2 = C_{\infty2}\frac{x_2}{x_{02}}exp\left(\frac{\alpha_2}{r}\right) - C_{\infty2} \approx C^*_2 x_{02}\left[\frac{x_2}{x_{02}}\left(1 + \frac{\alpha_2}{r}\right) - 1\right] \tag{50}$$

Here the exponents were expanded in series. We now can solve the coupled differential equations 1, 2, 49 and 50, calculating the volume flows for the Components 1, 2 and evaluating the change in the particle size and composition vs. time.
Figure 4 shows such dependence for a particular set of parameters. We chose the physical values that would be close to a scenario of dissolution of a d = 1 μm spherical emulsion drop made with two oils that differ in solubility in water by a factor of 1000, interfacial tension of 20 dyn/cm and initial molar fraction of the less soluble oil of 0.1. We can see from the graph that after a short rapid initial change in the particle size (pre-lock-in), the kinetics enters the almost perfectly linear stage, which lasts nearly to the end of the droplet existence. At the same time, the molar fraction of the less soluble oil quickly increases during pre-lock-in and then stays nearly constant, till it sharply increases in the very end (late-lock in).

The calculations of this example are based on the parameter set such that the Raoult effect overweighs the Laplace effect initially: $L_1 = 15.6$. However, as the droplet dissolves, eventually, it will need to pass through the stage when the effects are comparable and then when the Laplace effect is larger than the Raoult effect and the dependence must become quadratic (or, in cubic coordinates of the graph of Figure 4, the line should become concave). It does happen for a short period of time in the very end and does not contribute to the overall lifetime of the drop in this case.

This argument is clearer from the next example. Figure 5 shows the same graphs for the 10 x larger interfacial tension, 200 dyn/cm, for which $L_1 = 1.56$ . The trends are similar but there are notable differences. The pre-lock-in stage is now much longer (in relative terms). Also, the graph of the particle volume as a function of time is less linear, showing a late-lock-in concave deviations from the linearity in the end.

This is even more so when the interfacial tension is increased slightly further to the (hypothetical value of) $\sigma$= 311 dyn/cm, when the $L_1$ =1 (Figure 6). Conversely, if the interfacial tension is decreased to 2 dyn/cm, when $L_1$=156, the graph shows nearly perfect linearity and the pre-lock in and late lock in parts are barely visible (Figure 7).

Table 1 provides the summary of the degree of linearity of the volume vs time plots and the degree agreement between the theoretical slopes as determined by eqn 27, and the numerical simulations for different values of the lock-in parameters.

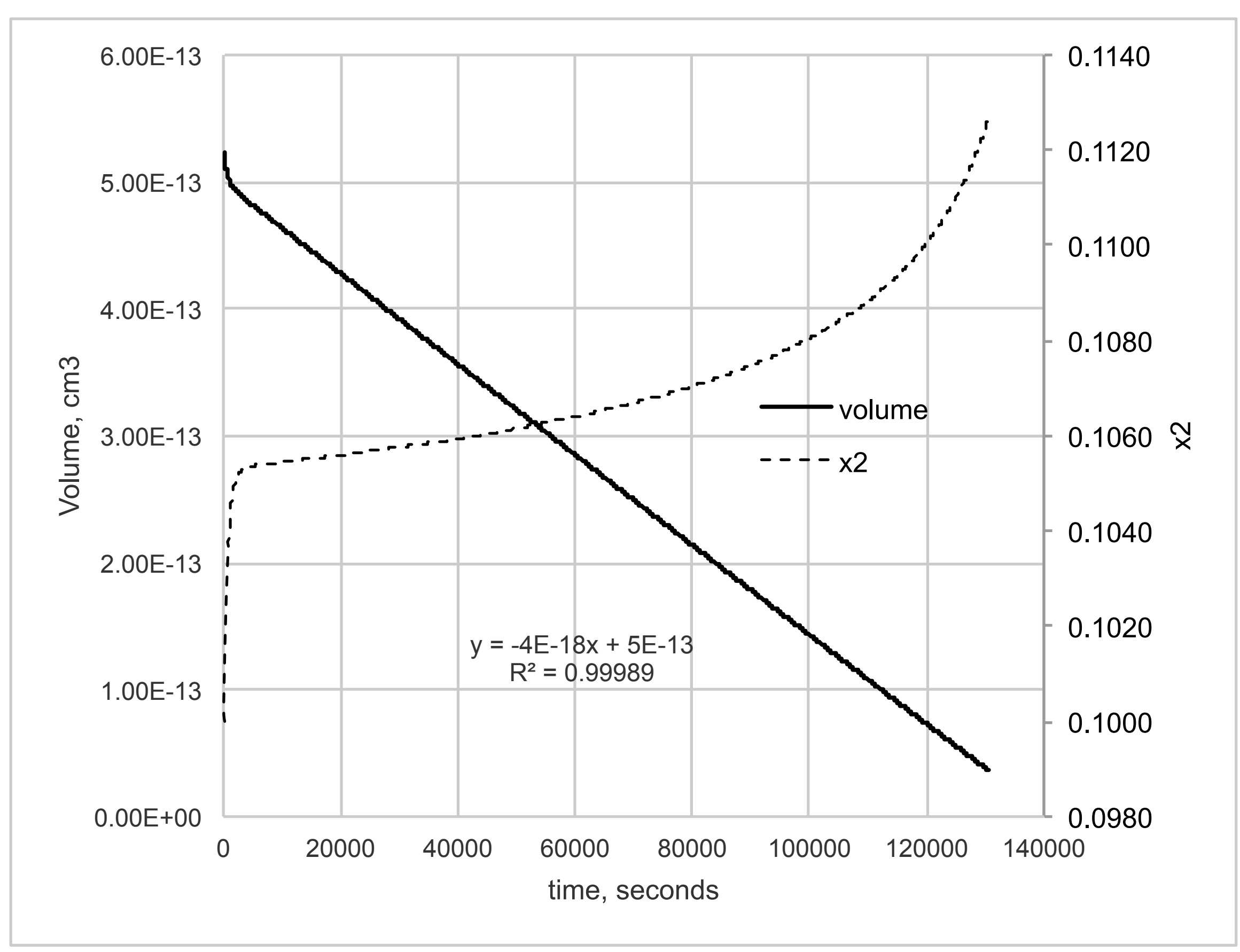

Figure 4. Kinetics of particle dissolution (bold line, left axis) and the changes in the concentration of the second component with time (dashed line, right axis). Parameters used: $C_1 = 10^{-6}$; $C_2=10^{-8}$; $D_1=D_2 = 10^{-5}$ $cm^2/s$; $\sigma$ = 20 dyn/cm, $V_{m1}= V_{m2}$ = 200 $cm^3/mol$; T= 300 K; $r_0$= 5 $10^{-5}$ cm, $\phi_{02}=x_{02}=0.1$; $L_1 = 15.6$.

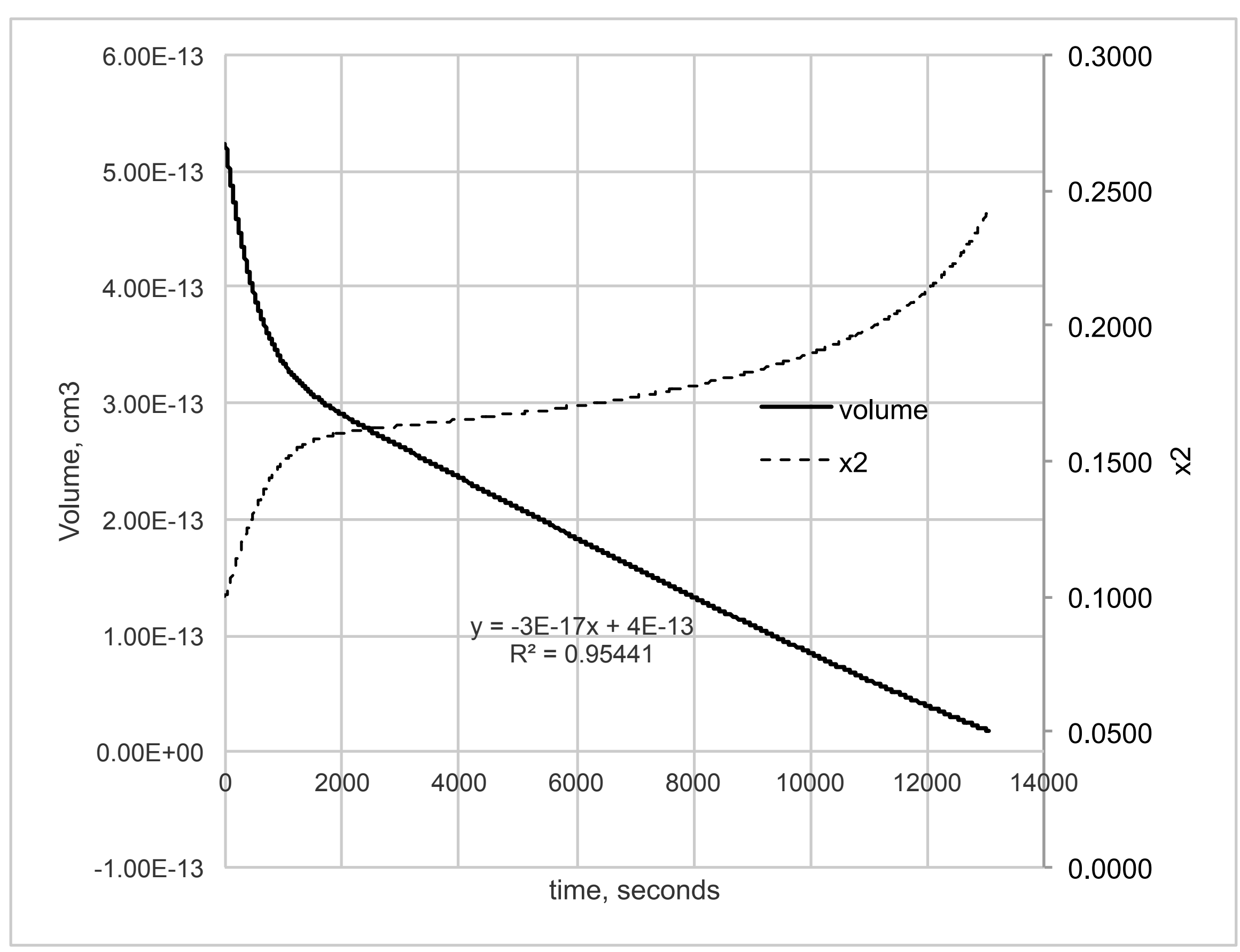

Figure 5. The same parameters as in Figure 3, except for the interfacial tension, $\sigma$= 200 dyn/cm; $L_1$= 1.56 .

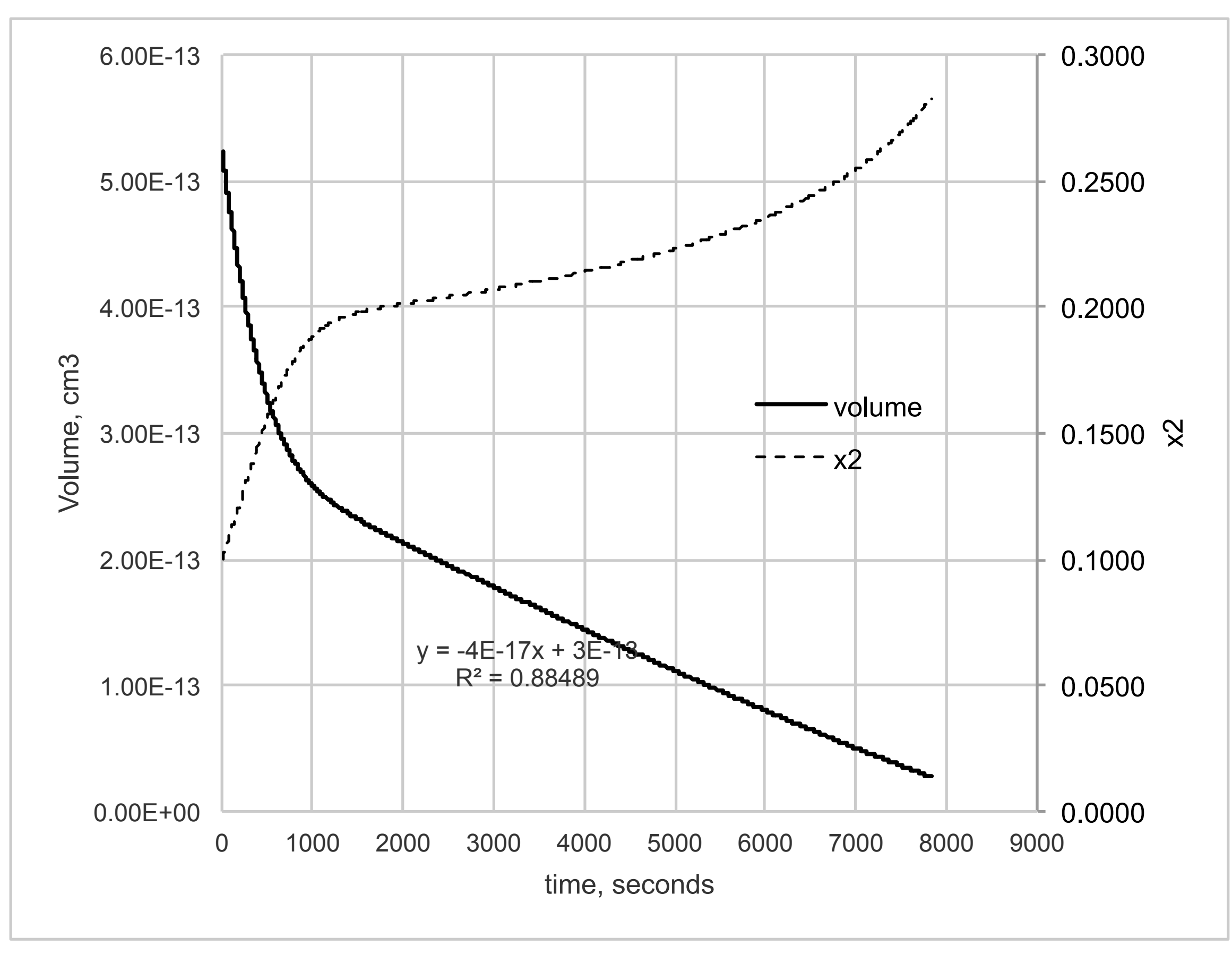

Figure 6. The same parameters as in Figure 3, except for the interfacial tension, $\sigma$= 311 dyn/cm; $L_1$= 1.

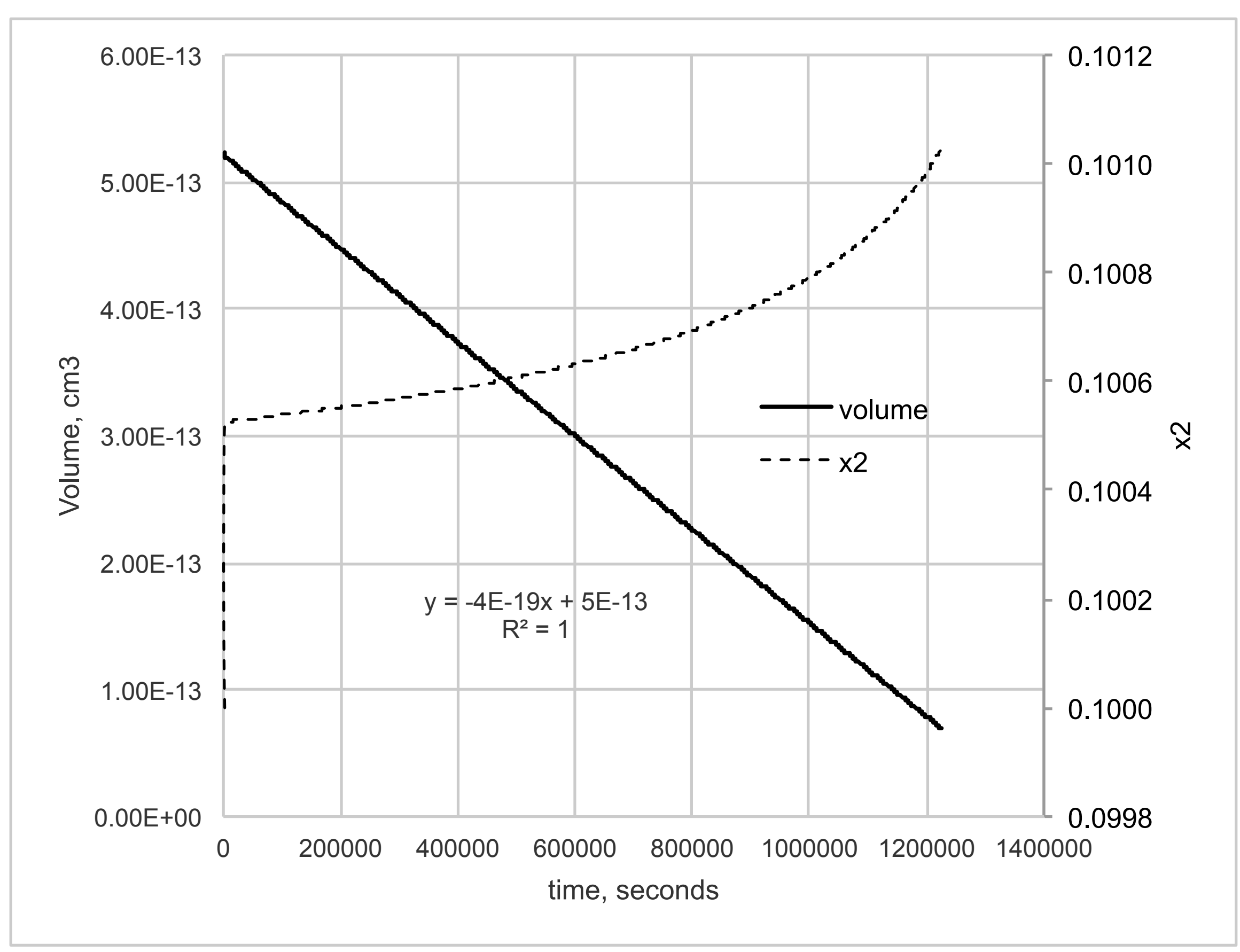

Figure 7. The same parameters as in Figure 3, except for the interfacial tension, $\sigma$= 2 dyn/cm; $L_1$= 156.

Table 1. Linearity of volume vs time dependence for different values of the lock-in parameters, and comparison of the V vs *t* slope with theory, eqn 27. Other parameters of the simulation: $C_1 = 10^{-5}$; $C_2=10^{-8}$; $D_1=D_2 = 10^{-5}$ cm$^2$/s; $V_{m1}= V_{m2} = 200$ cm$^3$/mol; $T$= 300 K; $r_0$= 5 $10^{-5}$ cm, $x_{02}= \phi_{02} = 0.1$.

| $\sigma$, dyn/cm | $L_1$ | $L_2$ | slope, cm$^3$/s | correlation coefficient of linear dependence | ratio: slope/ eqn 27 |
|---|---|---|---|---|---|
| 0.2 | 1560 | 111 | 4.00 $10^{-20}$ | 1.000 | 0.99 |
| 2 | 156 | 111 | 3.98 $10^{-19}$ | 1.000 | 0.99 |
| 20 | 15.6 | 111 | 3.81 $10^{-18}$ | 0.9999 | 0.95 |
| 200 | 1.56 | 111 | 2.58 $10^{-17}$ | 0.99 | 0.64 |
| 311 | 1 | 111 | 3.36 $10^{-17}$ | 0.98 | 0.53 |

## Diffusion of the sparingly soluble component

In the next round of numerical simulations, we explored the effect of the finite solubility of the additive on the dissolution rate. To do this, we chose the interfacial tension that is much smaller than in the previous simulation, $\sigma$= 0.2 dyn/cm and the same initial diameter of 1 μm. This allowed to have the Raoult effect larger than the Laplace effect by a significant margin, $L_1 >> 1$. However, the stabilizing effect still could be reduced because of the slow diffusion of the trapped species from the particle.
The results of the simulations over a large range of the initial volume fractions of the insoluble additive, from $\phi_{02}$ = $10^{-4}$ to 0.5 with the $L_2$ value varied from 100 to 0.01; we also varied the molar volume of components, with $V_{m1}$ and $V_{m2}$ at 200 and 200, 100 and 400, 400 and 100, 400 and 50, and 50 and 400 of $cm^3$/mol, respectively. The solubilities of the components were set at $10^{-6}$ and $10^{-8}$. The diffusion coefficients were set at $10^{-5}$ $cm^2$/s.
The linearity of the particle volume with time was observed in all cases, including the cases when $L_2$ was small and a strong bleeding of the second component was seen; in other words, the lack of the stabilizing action was seen not as the deviations from the linearity, but as the lack of the effect on the dissolution rate. This is not surprising as at low $L_2$ values, the kinetics is expected to evolve into one of the pure first component, which is also cubic.
Figure 8 compares predictions of the extrapolatory equation 31 with the numerical simulations for various values of molar volumes of the components. As expected, the agreement is nearly perfect for the volume fractions $\phi_{02}$ of 0.1 and larger, where 'exact' eqn 27 is expected to hold. One can see that the extrapolatory part of the equation also functions quite well, within about 20% deviations, even for very asymmetric molar volumes, differing by the factor of 8. We need to emphasize, that the equation (36) proposed in our previous work , ref [2], does not handle well the case of asymmetric molar volumes, and the deviations between the equation (36) and numerical simulations as large as a factor of 5 would have been seen.
It is of interest to understand the mechanism of how the stabilization mechanism breaks down when the insoluble component shows a considerable bleeding. Figure 9 shows the increase of $x_2$ at the stage of the pre-lock-in plotted versus its initial value $x_{02}$. The graph compares the numerically calculated values with the prediction of eqn 41. The agreement is quite good at large values of $x_2$ till the value of $L_2$ reaches about 1; at lower values of $L_2$ there is still some increase in $x_2$, but it is not sufficient for the complete lock-in; that is, the insoluble component still concentrates somewhat, but not to the level necessary to bring the complete lock-in.

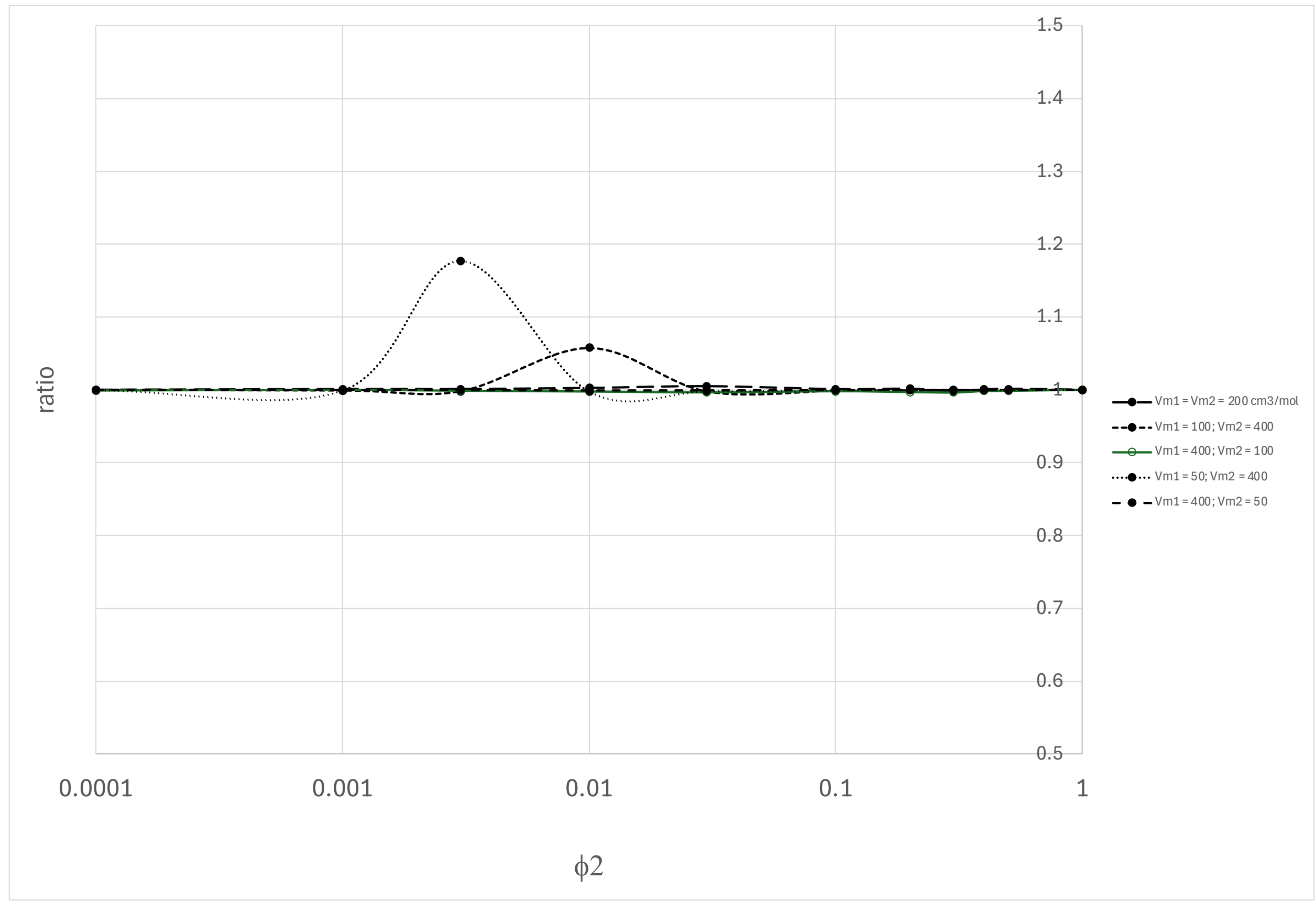

Figure 8. Ratio of the numerically calculated dissolution rate to the one predicted by eqn 31, for different sets of the molar volumes of the components shown on the graph. $D_1 = D_2 = 10^{-5}$ cm$^2$/s; $\sigma$= 0.2 dyn/cm ; $d_0$ = 1 μm, $C^*_{\infty 1} = 10^{-6}$ ; $C^*_{\infty 2} = 10^{-8}$, T = 300 K

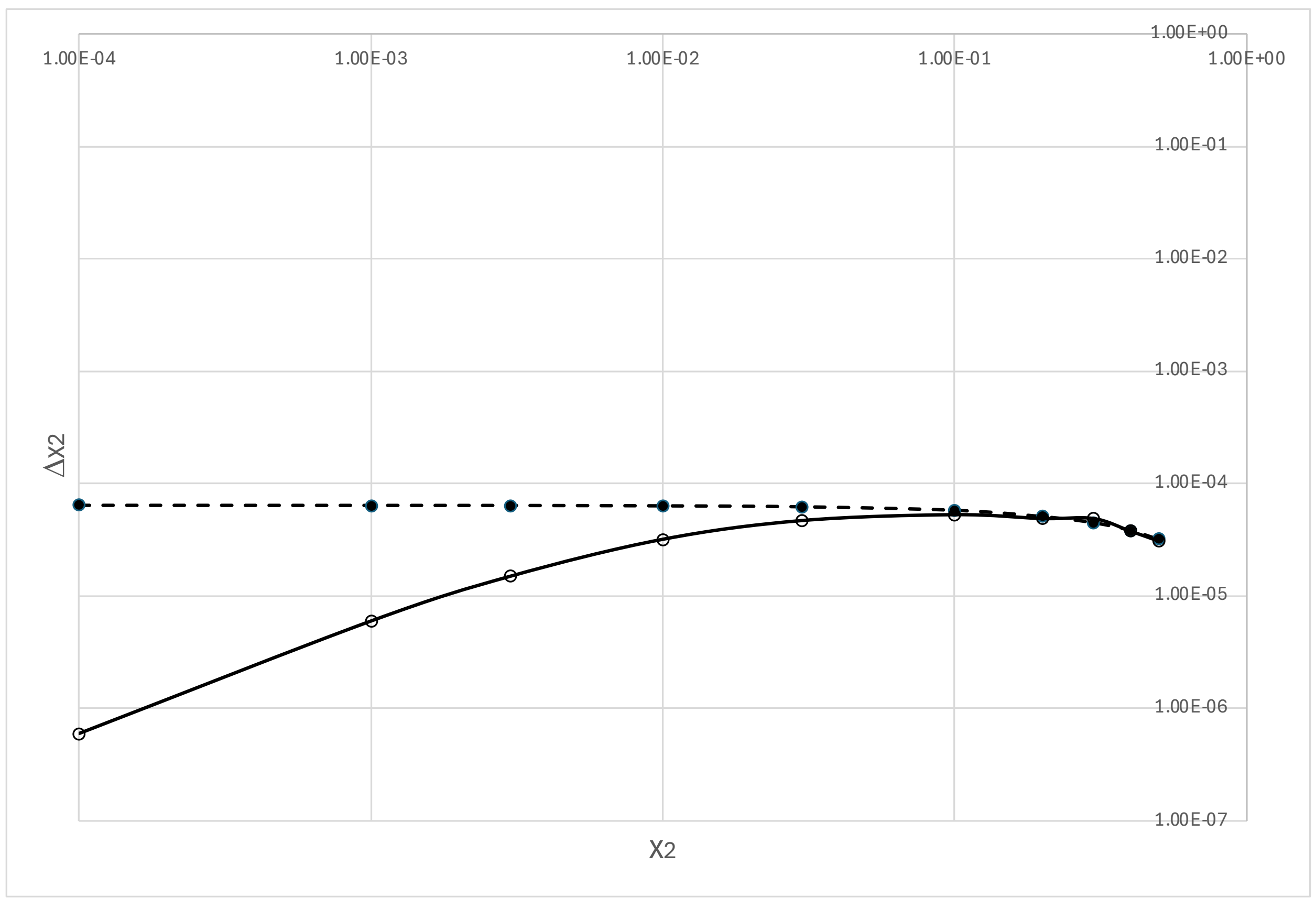

Figure 9. Increase in molar fraction of the second component during pre-lock in. Dashed line: eqn 41, under assumption of zero bleeding; full line, numerical simulations at $V_{m1}$ = $V_{m2}$ = 200 cm$^3$/mol; the rest of parameters are the same as in Figure 8. A complete lock-in is observed at $\phi_{02}$=$x_{02}$ larger than approximately 0.1; at smaller values of $x_{02}$, the lock-in is partial due to bleeding of the second component.

## Conclusions

The problem of dissolution of a two-component spherical particle onto a macrophase is studied. The driving force of the dissolution is the excess Laplace pressure inside the particle due to the surface tension effect. Three stages of dissolution are identified. In the first stage, called pre-lock-in, the concentration of the poorly soluble component undergoes a quick increase, and the system enters the lock-in state, in which the Laplace pressure effect on the chemical potential of the more soluble component is nearly completely counterbalanced by the Raoult effect. After this, the dissolution kinetics enters a steady state called the lock-in mode. In the process, the concentration of the sparingly soluble component continues to increase, first slowly and then more rapidly in the very end of the particle lifetime, when the relative changes in concentration become significant; the latter stage is called the late lock-in mode. If the initial concentration of the poorly soluble component is above a certain threshold, that is , the first lock-in number $L_1$>>1, the dissolution kinetics still nearly exactly follows the classical cubic law. The extrapolatory equation 31 for the rate of dissolution is derived; it is extension over our previous formula, introduced in ref [2]. The equation joins together the dissolution regimes over the whole range of compositions , from $x_{02}$ = 0 to 1, and accounts for the molar volume difference of the disperse phase components and for the solution nonideality.
The model can be substantially leveraged by Ostwald ripening theory [6,7], as it has been argued by us earlier [2]. The case of $L_1$ >>1 is in fact directly leverageable due to a simple scaling argument, and the ripening rate suppose to follow eqn 30, whereas the particle size distribution to follow the same pattern as the classical Lifshits-Slezov-Wagner theory [6, 7]. This will be discussed in the follow-up publication in more detail. The case of $L_1$~<1 is however not leverageable directly; a bimodal particle size distribution is predicted to emerge, with the smaller particle size fraction enriched by the poorly soluble component.

**Acknowledgment**

Comments of Prof. Håkan Wennerström on the manuscript are much appreciated.

**References**

1. W.I. Higuchi, J. Misra, Physical Degradation of Emulsions Via the Molecular Diffusion Route and the Possible Prevention Thereof, Journal of Pharmaceutical Sciences, Volume 51, Issue 5,1962, Pages 459-466
2. A.S. Kabal'nov, A.V. Pertzov, E.D. Shchukin, Ostwald ripening in two-component disperse phase systems: Application to emulsion stability, Colloids and Surfaces, Volume 24, Issue 1,1987, Pages 19-32.
3. A. J. Webster and M. E. Cates, Stabilization of Emulsions by Trapped Species, Langmuir 1998 14 (8), 2068-2079
4. Crank, John. *The mathematics of diffusion*. Oxford university press, 1979.
5. Bodenstein, M. "Eine theorie der photochemischen reaktionsgeschwindigkeiten." Zeitschrift für physikalische Chemie 85.1 (1913): 329-397.

6. I. M. Lifshitz and V. V. Slezov, “On diffusional decay kinetics of supersuturated solid solution,” Zh. Eksp. Teor. Fiz. 35, 497–505 (1958)
7. C. Wagner, “Theorie der Alterung von Niderschlagen durch Umlösen (Ostwald Reifung),” Z. Electrochem. 65, 581–591 (1961)